\documentclass[usenatbib]{mn2e}
\usepackage{epsf}

\makeatletter
\def\Mpc{\mathop{\mathrm{Mpc}}\nolimits}
\def\kpc{\mathop{\mathrm{kpc}}\nolimits}
\def\cm{\mathop{\mathrm{cm}}\nolimits}
\def\km{\mathop{\mathrm{km}}\nolimits}
\def\sec{\mathop{\mathrm{s}}\nolimits}
\def\yr{\mathop{\mathrm{yr}}\nolimits}
\def\Gyr{\mathop{\mathrm{Gyr}}\nolimits}
\def\Kelvin{\mathop{\mathrm{K}}\nolimits}
\def\keV{\mathop{\mathrm{keV}}\nolimits}
\makeatother

\newcommand{\be}{\begin{equation}}
\newcommand{\ee}{\end{equation}}

\newcommand{\eps}{\varepsilon}
\newcommand{\OM}{\Omega_M}
\newcommand{\Ob}{\Omega_b}
\newcommand{\OL}{\Omega_\Lambda}
\newcommand{\FeH}{[\mathrm{Fe}/\mathrm{H}]}
\newcommand{\ts}{t_\ast}
\newcommand{\tdyn}{t_\mathrm{dyn}}

\newcommand{\rhogas}{\rho_\mathrm{gas}}
\newcommand{\zini}{z_\mathrm{ini}}
\newcommand{\fb}{f_\mathrm{b}}
\newcommand{\mDM}{m_\mathrm{DM}}
\newcommand{\mb}{m_\mathrm{b}}
\newcommand{\Mclump}{M_\mathrm{cl}}
\newcommand{\Rclump}{R_\mathrm{cl}}
\newcommand{\dclump}{d_\mathrm{cl}}
\newcommand{\Mstar}{M_\ast}
\newcommand{\Mcg}{M_\mathrm{cg}}
\newcommand{\Vc}{V_\mathrm{c}}
\newcommand{\jtot}{j_\mathrm{tot}}

\newcommand{\jstar}{j_\ast}
\newcommand{\jnormstar}{\tilde{\jmath}_\ast}

\title[High-$z$ Galaxy Formation with WDM]
{Inhomogenous Primordial Baryon Distributions on
Sub-Galactic Scales: High-$z$ Galaxy Formation with WDM}
\author[J. Sommer-Larsen, P. Naselsky, I. Novikov and M. G\"{o}tz,]{Jesper
    Sommer-Larsen\thanks{E-mail: jslarsen@tac.dk}, Pavel Naselsky, Igor Novikov
    and Martin G\"{o}tz\\
    Theoretical Astrophysics Center,
    Juliane Maries Vej 30, 2100 Copenhagen, Denmark}

\begin{document}
\date{Accepted ---. Received ---; in original form 2003 September 11}

\pagerange{\pageref{firstpage}--\pageref{lastpage}} \pubyear{2003}

\maketitle

\label{firstpage}

\begin{abstract}
For the Warm Dark Matter (WDM) cosmological model the implications of strongly 
inhomogenous, primordial baryon distribution on sub-galactic scales 
for Big Bang Nucleosynthesis,
Cosmic Microwave Background anisotropies and Galaxy Formation (including
fully non-linear evolution to $z$=0) are discussed, and the inflationary
theory leading to such distributions is briefly reviewed. It is found that
Big Bang Nucleosynthesis is essentially unaffected relative to SBBN and
that the change in recombination history at $z~\sim~1500-700$ relative to
``standard'' theory leads to differences in the anisotropy and polarization
power spectra, which should be detectable by the Planck satellite provided
systematic effects can be accounted for. Moreover, it is shown by fully
cosmological, hydro/gravity simulations that the formation of galactic
discs is only weakly affected by going from smooth to highly non-homogenous, 
initial baryon
distributions. In particular, the final disc angular momenta at $z$=0
are as large as for the standard case and the ``disc angular momentum
problem'' is solved to within a factor of two or better without invoking
(hypothetical) energetic feedback events.  A very desirable
difference relative to the the standard WDM model, however, is that the on-set
of star (and AGN) formation happens
earlier. For the ``optimal'' free-streaming
mass scale of $M_f~\sim~1.5~\cdot~10^{11}~h^{-1}~M_{\odot}$ the redshift
of formation of the first stars increases from $z_\mathrm{*,i}$=4-5 to 
$\ga$6.5, in much better agreement with observational data on high-redshift 
galaxies and QSOs. It will, however, not be possible to push $z_\mathrm{*,i}$
above $\approx$10, because at higher redshifts the gas velocity field is
nowhere compressive. Probing the ``dark ages'' will hence enable a
direct test of this theory.

\end{abstract}

\begin{keywords}
galaxies: formation --- cosmology: theory --- cosmology: dark matter --- 
cosmic microwave background
\end{keywords}

\section{Introduction}

One of the most intriguing challenges in modern cosmology is
determining the origin and properties of the dark matter in the Universe.
Progress can be made either through attempts to detect dark matter
(elementary particles, massive compact halo objects (MACHOs), primordial
black holes etc.) or indirectly by studying the implications of various dark 
matter
candidates for the formation of galaxies, clusters of galaxies and the
large-scale structure. The former approach has so far not given any definite
results (except for the possible detection of micro-lensing by 
$\sim$0.5$M_\odot$ MACHOs in the Galactic halo), so one is currently left 
with the latter. 

The salient feature about ``conventional'' cold dark matter (CDM)
is that as long as the dark matter particles are much heavier than 1 keV, 
the actual particle mass does not matter for structure formation (assuming
that the dark matter is elementary particles --- note though,
that axions, despite being ultra-light, behave like CDM). Structure
formation in the CDM scenario faces a number of well known problems on
galactic scales:
1) the steep central cusps problem (e.g. Moore et al. 1999b), 
2) the angular momentum problem (e.g. Sommer-Larsen, G\"otz \& 
Portinari 2003), 3) the missing satellites problem (e.g. 
Klypin et al. 1999) and possibly 4) the disc heating problem (e.g. Moore et al.
1999a). Going to the warm dark matter (WDM) structure 
formation scenario can considerably improve on problems 2-4 and possibly also 
on problem 1. On much larger scales, the scarcity of galaxies in voids could be
a potential problem for CDM \citep{P01}, which also may be remedied by going
to WDM \citep{BOT}.

The WDM structure formation scenario has, however, also at least one major
problem in that it is difficult to get early enough galaxy formation to
match observational constraints from high-$z$ galaxies and QSOs. For a
free-streaming mass scale of $M_f~\sim~1.5~\cdot~10^{11}~h^{-1}~M_{\odot}$,
which \citet{Gsld} found was the optimal value for solving the angular
momentum problem, the first star formation takes place at $z\la6$ (see section
5). In contrast, both galaxies and QSOs have now been observationally detected
at $z>6$ (e.g., Hu et al. 2002, Fan et al. 2001) and the recent WMAP results 
possibly indicate that the reionization of the Universe took place at 
$z_\mathrm{re}\ga10$ \citep{S.03}.

The implicit assumption in standard WDM galaxy formation simulations is
that at early times the baryons and the dark matter are distributed in the
same way. The possibility exists, however, that the baryons could have
a strongly inhomogeneous initial (primordial) density distribution. If on the
(small) scale of these ``baryon'' clouds the fluctuations in the total matter 
density (baryons, dark matter and radiation, including neutrinoes) are 
isocurvature fluctuations
then the baryon clouds will not be smoothed out by Silk damping, nor will
they suffer growth by gravitational instability any faster than the 
fluctuations in the dark matter density \citep{NN}.
Because of the higher gas density
in the baryon clouds, the clouds will however cool out and form stars
earlier than is the case in standard WDM simulations. The purpose of this
paper is to quantify the effects of such initially inhomogeneous baryon
distributions on the epoch of first star/galaxy formation as well as
the subsequent disc formation. We also calculate how Big Bang Nucleosynthesis 
(BBN) and Cosmic Microwave Radiation Background (CMB) anisotropies are 
affected by going to this scenario, which we shall denote the WDMB model
in the following.
 
In section~2 we describe the mix between adiabatic and isocurvature 
perturbations in the WDMB model, in section~3 inflationary models
leading to such initial baryon clouds and in section~4 the resulting BBN and
CMB anisotropy and polarization for the WDMB model. In section~5 we present
our fully cosmological, hydro/gravity disc galaxy formation simulations, and
finally, in section~6, we summarize our conclusions.

\section{The Mix between Adiabatic and Isocurvature Perturbations in the
WDMB Model}

In the framework of modern theories of inflation
there are a lot of possibilities for having a mix of perturbation modes
in the composite fluid, which are discussed by \citet{RL},
\citet{BMR}, \citet{PS},
\citet{AF}, \citet{BMT} and others.
The general
idea for classifying modes of perturbations is based on a
simple definition of the isocurvature modes. They do not perturb the
gravitational potential meaning that the initial fluctuations of the total
matter density $\rho_{\mathrm{tot}}$ are zero.
In terms of the total density perturbation $\delta\rho_{\mathrm{tot}}$ this
corresponds to
\begin{equation}
\delta\rho_{\mathrm{tot}}=\sum\limits_{i=0}^{N}\rho_{i}\delta_i +
4\rho_{\gamma}(1+R_{\nu \gamma})
\frac{\delta T}{T}=0,
\label{eq1}
\end{equation}
where $\rho_{i}$ denotes the density of each massive species including
baryons and different kinds of CDM and WDM particles,
$\delta_i = \delta\rho_{i}/\rho_{i}$ is the density contrast
for each massive component, $R_{\nu \gamma}$ is the density ratio
between neutrinos $\rho_{\nu}$ and black body radiation
$\rho_{\gamma}$, and $\delta T/T$ are the CMB temperature
perturbations.

Using  Eq.~(\ref{eq1}) one can find a certain peculiar mode
(or modes) which compensates the contribution of baryonic perturbations
to $\delta\rho_{\mathrm{tot}}$, i.e.\ it corresponds to the condition
$\rho_\mathrm{b}\delta_\mathrm{b}=-\rho_{dm}\delta_{dm}$ (we assume here
without loss of generality that there is just one main dark matter component).
We will denote
this mode a compensating isocurvature mode (CIM) for the
dark matter \citep{NN}. As one can see this
mode corresponds to $\delta T/T=0$. This means that the CIM
are equivalent to an isotemperature perturbation. Note that exactly the same mode was described by \citet{AF}, but for
compensation between quintessence scalar field perturbations and some
of the CDM particles.

The new aspect in the evolution of the isocurvature perturbations comes from the
WDM model. If the rest mass of the dark matter particle
is comparable to $m_x\simeq 1$ keV, then practically all perturbations at
spatial scales smaller than the free streaming scale
$\lambda_{\mathrm{fs}}\simeq
0.2\left(\frac{g_{\mathrm{WDM,dec}}}{100}\right)^{-1/3}
\left(\frac{m_x}{\mathrm{keV}}\right)^{-1}$ Mpc are damped while at scales
above $\lambda_{\mathrm{fs}}$
they behave like in the CDM model. The corresponding free-streaming mass
in terms of the scale
$\lambda_{\mathrm{fs}}$ is \citep{Gsld}
\begin{equation}
M_{\mathrm{fs}}=3.7~10^{11}~\Omega_{\mathrm{WDM}}~h^{-1}
\left(\frac{\lambda_{\mathrm{fs}}}{0.1
h^{-1} \Mpc}\right)^3 M_{\odot}
\label{eq2}
\end{equation}
where $h= H_0/(100 \mathrm{km/s/Mpc})$ is the Hubble constant,
$g_{\mathrm{WDM,dec}}$ is
the effective number of particle degrees of freedom when the WDM particle(s)
decouple from the cosmic plasma, and
\begin{equation}
\Omega_{\mathrm{WDM}}h^{2}=
\left(\frac{g_{\mathrm{WDM,dec}}}{100}\right)^{-1}
\left(\frac{m_x}{\mathrm{keV}}\right).
\label{eq3}
\end{equation}

This means that for pure adiabatic initial perturbations in the WDM model
all inhomogeneities at the scales $M\le M_{\mathrm{fs}}$ are damped and
baryons and WDM particles
follow an approximately homogeneous and isotropic spatial distribution.

The situation changes dramatically if we instead of pure adiabatic
perturbations have 
a mix between adiabatic and isocurvature modes of fluctuation. Let us
discuss a model with isocurvature baryonic perturbations which have a typical
spatial scale $\lambda\ll \lambda_{\mathrm{fs}}$, but the corresponding
amplitudes of perturbation
are much higher than for the adiabatic tail of perturbations. This model
is close to the baryon isocurvature fluctuations model \citep{DS}
and manifests itself as a combination of an adiabatic WDM model plus small-scale
baryonic clouds. We shall denote this model the WDMB model.

Although we will consider in the non-linear galaxy formation calculations
(Section~5) a strongly in-homogeneous, initial baryon density field + CIM,
within the framework of isocurvature modes even if the
the condition $\rho_\mathrm{b}\delta_\mathrm{b}=-\rho_{dm}\delta_{dm}$ is
{\it not} satisfied (e.g., with initially strongly non-homogenous baryon
+ smooth dark matter density fields) the baryon density field will neither
be smoothed out due to Silk damping, nor grow non-linear much before
the dark matter density field due to gravitational instability (e.g. 
Naselsky \& Novikov 2002). We note, however, that for baryon ``cloud'' masses 
$\Mclump \la M_\mathrm{diff} \simeq 10-20~M_{\odot}$ the baryon density 
field will be smoothed before the epoch of recombination due to baryon
diffusion \citep{ZN}.

Finally we note that \citet{DD} recently also argued for small scale
peculiarities in the power spectrum, albeit on other grounds.

\section{Baryonic Clouds as a Relic of the Baryogenesis Epoch}

It is necessary to note that the idea of a non-homogeneous distribution
of baryonic matter at small scales is not new. The importance of
entropy perturbations in the history of the cosmological
expansion was {\it ad hoc} demonstrated by \citet{DZN}
and Peebles (1967,1994) and recently
generalized taking into account the multi-species structure of the
cosmological plasma by \citet{GO}, \citet{HL},
\citet{PJ}. The possible inhomogeneities in
the baryon distribution at the epoch of nucleosynthesis
(Inhomogeneous Big Bang Nucleosynthesis (IBBS)) has been widely discussed in
the literature (see the review by Jedamzik \& Rehm 2001) in connection
with the quark-hadron phase transition. But the typical scales of such
inhomogeneities are extremely small compared to the typical mass
scales of interest here - see below.
Other events or processes have been suggested as possible sources of
isocurvature perturbations partly connected with baryon re-distribution
in space. For example, cosmic strings and corresponding  currents
and magnetic fields could generate specific features in  baryonic
matter distribution \citep{MB}. \citet{YS}, \citet{DS}, \citet{PS}, 
\citet{NSM} have suggested various ways of generating 
isocurvature perturbations in the framework of inflation
theory. Recently \citet{NN} and \citet{DNNN}
have shown that baryonic clouds could mimic the acceleration of
hydrogen recombination at redshift $z\simeq 10^3$ and produce corresponding
peculiarities in the cosmic microwave background (CMB) anisotropy and
polarization power spectrum.

Probably, the most interesting question in
modern baryogenesis theory is how exactly the
primordial soup of particle species
became charge symmetry breaking  and what  the spatial distribution of the
baryons might be. We will consider one of the most attractive models for
baryonic cloud creation just after the end of inflation, suggested
by \citet{DS} and similar to the \citet{AD} baryogenesis model. We will 
assume that
two scalar fields $\Phi$ and $\phi$ play as crucial a role in inflation as in
charge symmetry breaking. Namely, that the field  $\Phi$  determines the 
general
properties of inflation while spontaneous CP violation is achieved by a
condensate
of the complex scalar field $\phi$. Following \citet{DS}, we will
describe
charge symmetry breaking using as the potential of the $\phi$ field
\begin{equation}
V(\phi)=m^2_{\mathrm{eff}}|\phi|^2 + \lambda
|\phi|^4\ln\left(\frac{|\phi|^2}{\sigma^2}\right),
\label{eq4}
\end{equation}
where
\begin{equation}
m^2_{\mathrm{eff}}=-m^2_0 + \beta T^2 +\mu (\Phi-\Phi_{\mathrm{cr}})^2
\label{eq5}
\end{equation}
is the effective mass of the  $\phi$ field, including interaction
with the inflaton $\Phi$.
$m_0$ is the vacuum mass of $\phi$, $\Phi_{\mathrm{cr}}$ the critical value
of the  $\Phi$ field when the
effective mass $m^2_{\mathrm{eff}}$ has
a global minimum, and the $\beta T^2$ term corresponds to the temperature
correction arising from the
interaction of  the $\phi$ field with the thermal bath, assuming
no conformal coupling between
$\phi$ and the curvature $R$. That all the
constants $m_0$, $\lambda$, $\mu$, $\Phi_{\mathrm{cr}}$
and $\beta$ are assumed to be free parameters of the
model allows us to optimize the resulting baryonic cloud properties.

Let us briefly describe some important stages in the  $\Phi$ and $\phi$
field evolution during and right after the
end of inflation. When inflation started, the initial value of the
$\Phi$  field is assumed to be
higher than the Planck mass, $\Phi_{\mathrm{in}}> m_{\mathrm{Pl}}$,
and higher than $\Phi_{\mathrm{cr}}$ as well,
and $m^2_{\mathrm{eff}}>0$. The potential $V(\phi)$ has
a global minimum at $\phi=0$ and
roughly $V(\phi)\propto m^2_{\mathrm{eff}}|\phi|^2$. As one can see, the charge
symmetry is unbroken. During the slow roll epoch of inflation the amplitude
of the  $\Phi$ field is a decreasing
function of time, which is typical, for example, for the model of chaotic
inflation,  and when $\Phi=\Phi_{\mathrm{cr}}$ the effective
mass of the  $\phi$ field reaches the minimum value
$ m^2_{\mathrm{eff,min}}=-m^2_0 + \beta T^2<0 $. The end of inflation
corresponds to the condition $\Phi\simeq 0 \ll \Phi_{\mathrm{cr}}$, while
$m^2_{\mathrm{eff}}=-m^2_0 + \beta T^2 +\mu \Phi^2_{\mathrm{cr}}>0$.
Obviously, in the vicinity of the point $\Phi=\Phi_{\mathrm{cr}}$ the
effective mass $m^2_{\mathrm{eff,min}}$ can be negative, or
positive but small, which means that it may happen that for some period of time
$|m_{\mathrm{eff,min}}|< H_{\mathrm{sr}}$, where
$H_{\mathrm{sr}}$ is the value of the Hubble parameter during the slow roll
decrease of
the  $\Phi$ field caused by
inflation. \citet{DS} have shown that the phase transition of the
$\phi$ field during the epoch
$|m_{\mathrm{eff,min}}|< H_{\mathrm{sr}}$ is essential
for the formation of bubbles, which are
governed by the quantum fluctuation of the  $\phi$ field. The corresponding
number density of the bubbles
as a function of their mass $M$ is \citep{DS}:
\begin{equation}
\frac{\mathrm{d}n(M)}{\mathrm{d}M} \simeq \frac{m^6_{\mathrm{Pl}}}{M^4_0}
\exp \left[ -\alpha -
\frac{\gamma}{4}\ln^2\left(\frac{M}{M_0}\right)\right]
\label{eq6}
\end{equation}
where
$$
M_0 = \frac{1}{4} m^2_{\mathrm{Pl}}H^{-1}_{\mathrm{sr}}\exp
\left[2 H_{\mathrm{sr}}(t_\mathrm{e} - t_{\mathrm{cr}})
-8\gamma^{-1}\right],
$$
and $\alpha=\delta + 16\gamma^{-1}$, $\delta=\frac{4\pi^2 m^2_0}{3\lambda
 H^2_{\mathrm{sr}}}$,
$$
\gamma=\frac{4\pi^2\mu}{3\lambda
 H^4_{\mathrm{sr}}}\left. \left(\frac{\mathrm{d}\Phi}{\mathrm{d}t}\right)^2
 \right|_{\Phi=\Phi_{\mathrm{cr}}}.
$$
$t_\mathrm{e}$ means
the end of inflation and $t_{\mathrm{cr}}$ is the moment when
$\Phi=\Phi_{\mathrm{cr}}$.
For the  $\Phi$ field using slow roll approximation we obtain
\begin{equation}
\dot{\Phi}\simeq -\frac{1}{3H_{\mathrm{sr}}}
\frac{\mathrm{d}V_{\mathrm{in}}(\Phi)}{\mathrm{d}\Phi},
\label{eq7}
\end{equation}
where $V_{\mathrm{in}}(\Phi)$ is the inflaton potential. That lead to the
following value for the
$\gamma$ parameter:
\begin{equation}
\gamma=\frac{4\pi^2\mu}{27\lambda
 H^6_{\mathrm{sr}}}\left. \left(\frac{\mathrm{d}V_{\mathrm{in}}(\Phi)}
 {\mathrm{d}\Phi}\right)^2\right|_{\Phi=\Phi_{\mathrm{cr}}}.
\label{eq8}
\end{equation}

As one can see from Eqs.~(\ref{eq6}) -- (\ref{eq8}), the distribution function
$\mathrm{d}n(M)/\mathrm{d}M$ is strongly peaked at
$M\sim M_0$, if the corresponding value of the $\gamma$ parameter
is  $\gamma\gg 1$. In this case
we can approximate  $\mathrm{d}n(M)/\mathrm{d}M\propto \delta(M-M_0)$, where
$\delta$
is the Dirac $\delta$-function, and the
corresponding mass $M_0$ is a function of the fundamental parameters, namely,
$H_{\mathrm{sr}}$, $t_\mathrm{e}$ and $t_{\mathrm{cr}}$.
\citet{DS} have shown that the simplest way to have a large value
for the baryon asymmetry inside
the bubbles is to assume that CP-nonconservation is proportional to the value
of the scalar field $\phi$ which is
much larger than that outside the clouds. Thus, because of bubble
formation during inflation, and
using CP-nonconserving effects, the baryonic matter can have some strong
perturbation strongly peaked
at some characteristic mass scale $M_0$. It is not possible at present to
calculate the value of $M_0$ from first principles, but we shall assume in
the following that $M_\mathrm{diff} \la M_0 \ll M_\mathrm{fs}$ (Section 2). 

For simplicity we suppose that all
baryonic clouds  have the same characteristic size
$R_{\mathrm{cl}}\propto M^{1/3}_0$, which
is much smaller than the size of the horizon $R_{\mathrm{rec}}$ at the
epoch of recombination ($z\sim 10^3$), $ R_{\mathrm{cl}}\ll R_{\mathrm{rec}}$.
We denote by $ \rho_{\mathrm{b,in}}$ and $\rho_{\mathrm{b,out}}$  the baryon
density inside and outside the clouds. Then the mean density
$\rho_{\mathrm{b,mean}}$ at
scales much greater than $R_{\mathrm{cl}}$ and distances between them, is
\begin{equation}
\rho_{\mathrm{b,mean}}=\rho_{\mathrm{b,in}} f + \rho_{\mathrm{b,out}}(1-f),
\label{eq9}
\end{equation}
where $f$ is the volume fraction of the clouds. We denote
\begin{equation}
\eta= \frac{\rho_{\mathrm{b,in}}}{\rho_{\mathrm{b,out}}}.
\label{eq10}
\end{equation}

Using this one obtains the following relations between the mean value of the
baryon density and its inner and outer values
\begin{equation}
\rho_{\mathrm{b,in}}=\frac{\eta \rho_{\mathrm{b,mean}}}{1+f(\eta -1)},
\label{eq11}
\end{equation}
and
\begin{equation}
\rho_{\mathrm{b,out}}=\frac{ \rho_{\mathrm{b,mean}}}{1+f(\eta -1)}.
\label{eq12}
\end{equation}
Using the quantities $f$ and $\eta$ we can define the baryonic mass fraction
of the clouds
\begin{equation}
F_\mathrm{b}=\frac{\eta f}{1+f(\eta -1)}.
\label{eq13}
\end{equation}

Obviously, all the parameters  $f$, $\eta$ and $F_\mathrm{b}$ are the result
of fine tuning the
inflaton $V_{\mathrm{in}}(\Phi)$ and $\phi$ scalar field potentials leading to
the formation of the baryonic
asymmetry in the Universe.

\section{Baryonic Clouds and their Cosmological Consequences}

There are several sources of information on $\Omega_\mathrm{b} =
\rho_{\mathrm{b}}/\rho_{\mathrm{cr}}$, where $\rho_{\mathrm{b}}$ and
$\rho_{\mathrm{cr}}$ are
the present values of the baryonic and critical densities.
One comes from the confrontation of the Standard Big Bang
Nucleosynthesis (SBBN) theory with observational data (see the review by
Fukugita, Hogan \& Peebles 1998). The corresponding value of the
baryonic density from this line of argument is
$\Omega_\mathrm{b} h^2=0.019 \pm 0.001$.
An additional empirical relation between baryonic and total matter fractions
$\fb=\Omega_\mathrm{b}/\Omega_\mathrm{M}\sim 0.1-0.15$ for $h=0.7$
comes from X-ray
data on clusters of galaxies \citep{Ca,Gef}.
For the most popular  $\Lambda$CDM cosmological model with
$\Omega_\mathrm{M}\simeq 0.3$ and $\Omega_{\Lambda} \simeq 0.7$, the
corresponding
value of $\Omega_\mathrm{b} h^2$ is $\sim 0.02$, in agreement with
the SBBN predictions \citep{Mason}.

An independent and important information about the baryonic fraction
in the Universe comes from recent CMB experiments such as
BOOMERANG \citep{dB}, MAXIMA-1 \citep{Han},
CBI \citep{Mason}, and DASI \citep{Kov}.
Fitting the CMB anisotropy power spectrum to the above-mentioned observational
data \citep{TZ,Wh2000,LP}
indicates that the baryon fraction parameter should be
significantly larger than the SBBN expected value, namely,  $\Omega_\mathrm{b}
h^2\simeq 0.03$. However, \citet{BC}
showed that new BOOMERANG, MAXIMA-1 and DASI data do not contradict
$\Omega_\mathrm{b} h^2=0.022 \pm 0.004$.

It is worth noting that the above mentioned methods for the baryonic
fraction density estimation from CMB and SBBN predictions are
based on the simple idea that the distribution of matter (including  dark
matter particles and baryons) is nearly homogeneous on all
scales, except for small fluctuations leading to galaxy and large-scale
structure formation. Typically, these are assumed to be adiabatic.
One can ask how sensitive conclusions based on the BBN + CMB data are to 
the presence of small-scale
baryonic (non-linear) clouds before cosmological recombination.

\subsection{Big Bang Nucleosynthesis (BBN)}

\begin{figure}
\epsfxsize=\columnwidth
\epsfbox{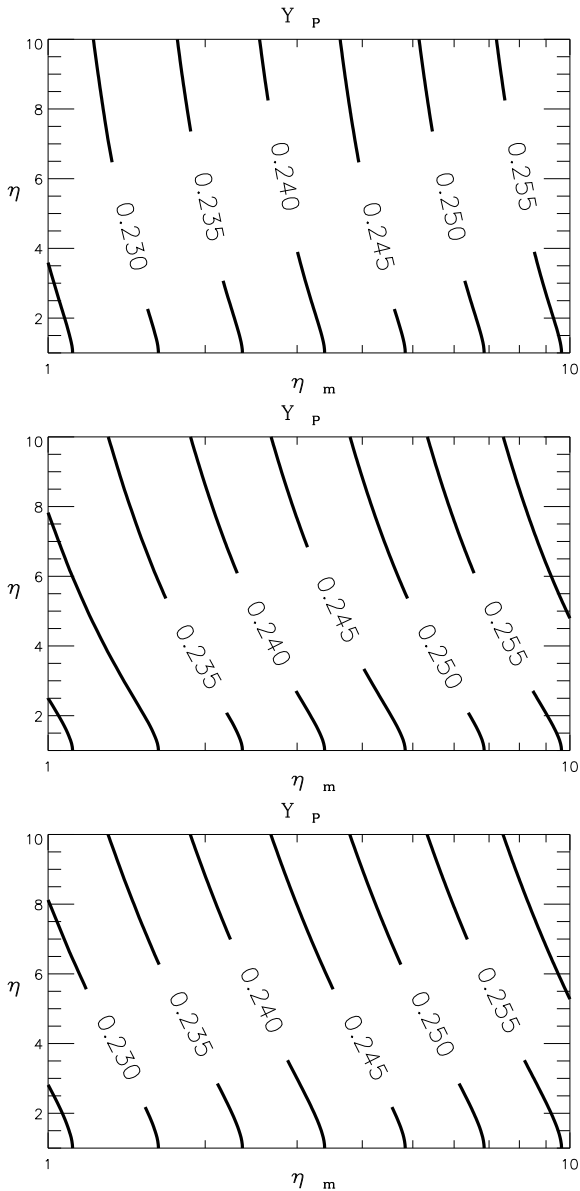}
\caption{The mean mass fraction of $^4$He as a function of $\eta_\mathrm{m}$ and
$\eta$
for different baryonic mass fractions $F_\mathrm{b}$. From top to bottom
$F_\mathrm{b}=0.2,0.5,0.8$.}
\label{fig1}
\end{figure}
\begin{figure}
\epsfxsize=\columnwidth
\epsfbox{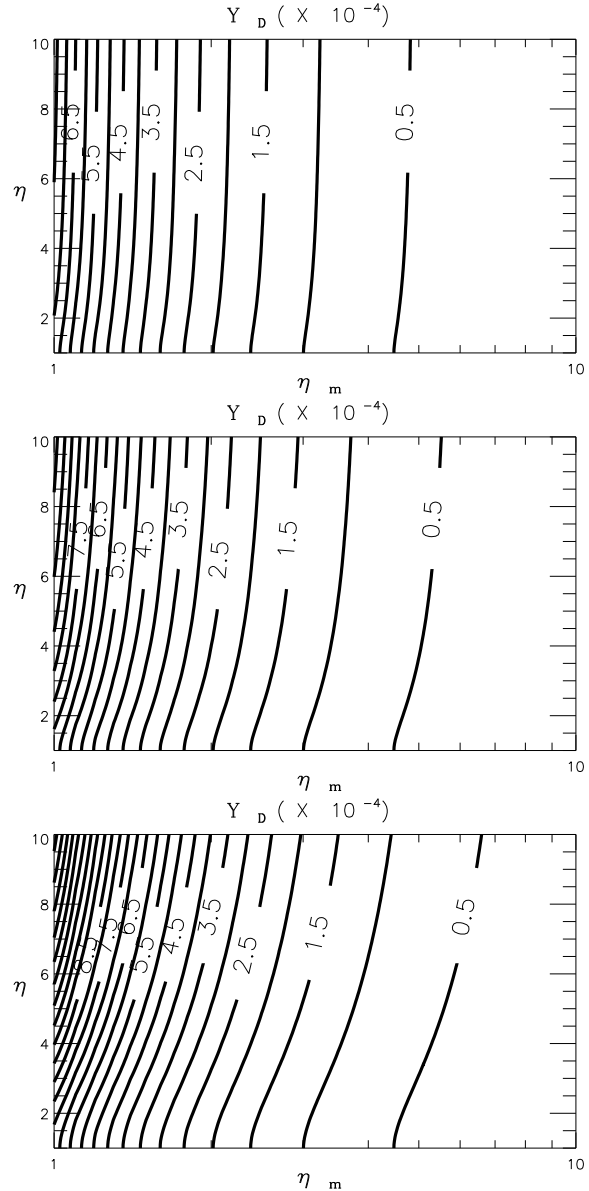}
\caption{The mean mass fraction of D as a function of $\eta_\mathrm{m}$ and
$\eta$
and the same  $F_\mathrm{b}$
parameters as in Fig.\ 1.}
\label{fig2}
\end{figure}

In the framework of the SBBN model the baryonic fraction of matter and
its
spatial distribution play a crucial role starting from the epoch, when the
balance between neutrinos ($\nu_\mathrm{e},\overline{\nu_\mathrm{e}}$),
neutrons ($n$) and
protons ($p$) in
reactions $n+\nu_\mathrm{e}\leftrightarrow p +e^{-}$,
$n + e^{+} \leftrightarrow p + \overline{\nu_\mathrm{e}}$,
$n\rightarrow p + e^{-} + \overline{\nu_\mathrm{e}}$
was broken. The corresponding time of violation of the neutrino-baryon
equilibrium is close to
$\tau_{\nu_\mathrm{e},p}\simeq 1$ sec when the temperature of the plasma was
close to $T_{\nu_\mathrm{e},p}\simeq 10^{10} K$ (see
the review by Olive, Steigman \& Walker 2000). The time scale
$\tau_{\nu_\mathrm{e},p}$ determines the
characteristic length $l_{\nu_\mathrm{e},p} \simeq c\tau_{\nu_\mathrm{e},p}$,
which in terms of the baryonic mass fraction
of matter corresponds to
\begin{equation}
M_{\nu_\mathrm{e},p}\sim m_{\mathrm{pl}}
\left(\frac{\tau_{\nu_\mathrm{e},p}}{t_{\mathrm{pl}}}\right)
\left. \left(\frac{\rho_\mathrm{b}}{\rho_{\gamma}}\right)
\right|_{t=\tau_{\nu_\mathrm{e},p}}
\hspace{-2mm} \simeq 
3~10^{-3}\left(\frac{\Omega_\mathrm{b}h^2}{0.02}\right) M_{\odot},
\label{eq14}
\end{equation}
where $t_{\mathrm{pl}}$ is the Planck time, $\rho_\mathrm{b}$ and
$\rho_{\gamma}$ are the densities of
baryons and radiation in the standard cosmological model without baryonic
clouds. The light elements
(chiefly, $^4$He and deuterium) were completely synthesized by 
$\tau_{\mathrm{end}}\sim 3\cdot 10^2 - 10^3$ s. In terms
of the baryonic mass scale it corresponds to
\begin{equation}
M_{\mathrm{end}}\simeq
M_{\nu_\mathrm{e},p}
\left(\frac{\tau_{\mathrm{end}}}{\tau_{\nu_\mathrm{e},p}}\right)^{3/2}
\hspace{-2mm} \simeq 
16~\left(\frac{\tau_{\mathrm{end}}}{300 s}\right)^{3/2}
\left(\frac{\Omega_\mathrm{b}h^2}{0.02}\right) M_{\odot}.
\label{eq15}
\end{equation}
Thus, if the characteristic mass scale $M_0$ for the baryonic clouds is  higher
than $M_{\mathrm{end}}$ ($\sim M_\mathrm{diff}$ cf. section 2),
the cosmological nucleosynthesis  in each cloud is independent of the others
and
the mean mass fraction of each chemical element is related with the mass
fraction of the clouds as
\[
\langle Y_k\rangle=\frac{1}{V}\int^{V_{\mathrm{cl}}}d^3
r\left(\frac{\rho_{\mathrm{in}}(\vec{r})}{\rho_{\mathrm{mean}}}\right)
Y^{\mathrm{in}}_k\left(\rho_{\mathrm{in}}(\vec{r})\right) +
\]
\begin{equation}
\frac{1}{V}\int^{V-V_{\mathrm{cl}}}d^3 r\left(
\frac{\rho_{\mathrm{out}}(\vec{r})}{\rho_{\mathrm{mean}}}\right)
Y^{\mathrm{out}}_k\left(\rho_{\mathrm{out}}(\vec{r})\right)
\label{eq16}
\end{equation}
where $V_{\mathrm{cl}}/V=f $, $V$ is the volume and $Y^{\mathrm{in}}_k$,
$Y^{\mathrm{out}}_k$ are the
inner and outer mass fractions of the
$k$th element.
If we assume that inside the clouds the density $\rho_{\mathrm{in}}(\vec{r})$
is uniformly and isotropically
distributed then from Eq.~(\ref{eq11})--(\ref{eq13}) and  Eq.~(\ref{eq16}) we
obtain
\begin{equation}
\langle Y_k\rangle=Y^{\mathrm{in}}_k(\rho_{\mathrm{in}})F_\mathrm{b} +
Y^{\mathrm{out}}_k(\rho_{\mathrm{out}})(1-F_\mathrm{b})
\label{eq17}
\end{equation}
In Fig.\ \ref{fig1} and Fig.\ \ref{fig2} we show the dependence
of the mean fraction of
$^4$He and D on
$\eta_\mathrm{m}=10^{10} \langle n_\mathrm{b}\rangle / n_\gamma=274
\langle \Omega_\mathrm{b}\rangle h^2$ and $\eta$
for such a model. The   three panels of Fig.\ \ref{fig1}
show different values of the
mass fraction $F_\mathrm{b}=0.2$ (top),
$F_\mathrm{b}=0.5$ (middle) and $F_\mathrm{b}=0.8$(bottom). As one can see
from Fig.\ \ref{fig1}, if the $\langle \Omega_\mathrm{b}\rangle h^2$ parameter
is close to $\langle \Omega_\mathrm{b}\rangle h^2=0.022$
($\eta_\mathrm{m} \sim 5$), and $\eta=1$,
then the mean mass fraction of helium is
equal to $Y_p$ in the standard model (without clouds) and corresponds to
$Y_p=0.247$.
For the model $F_\mathrm{b}=0.2$, $1\le \eta \le 10$ and 
$\eta_\mathrm{m} \sim 5$ the mean helium  mass fraction lies in the range
$0.247\le \langle Y_p\rangle\le 0.252$
which corresponds to the $2\%$ relative deviation from the value
$Y_p=0.247$. For the model with
$F_\mathrm{b}=0.5$, as it follows from Fig.\ \ref{fig1}, the maximum value of
$\langle Y_p\rangle=0.256$ is achieved for $\eta=10$ and the
corresponding deviation is $3\%$. For the model with
$F_\mathrm{b}=0.8$ we obtain a
quite similar result:
$\langle Y_p\rangle\le 0.256$ for all values in $\eta\le 10$ . 
Since the uncertainty in the observationally determined value of $Y_p$ is
$\sim$~5\% (when systematics are included --- B.~Pagel 2003, private 
communication) then even if $\eta_\mathrm{m}$ was known to high precision
(e.g. from the CMB data) it would not be possible to distinguish between
the smooth and clumpy initial baryon distribution scenarios using BBN.

For the primordial mass fraction of deuterium the dependence of
$\langle Y_\mathrm{D}\rangle$
on  $\eta$,  $\eta_\mathrm{m}$ and
$F_\mathrm{b}$ is shown in Fig.\ \ref{fig2}. As can be seen from
the figure, for
$\eta_\mathrm{m} \sim 5$ and $\eta=1$ the
mass fraction of deuterium is close to
$\langle Y_{\mathrm{D,st}}\rangle\simeq 4\cdot
10^{-5}$. Moreover, for $\eta\le 10$ and
$F_\mathrm{b}=0.2,0.5,0.8$ the corresponding variation of
$\langle Y_\mathrm{D}\rangle$ is $\la$ $20-30\%$ for $\eta_\mathrm{m} \sim 5$.
As this is less than the uncertainty on the observationally determined value
of $\langle Y_\mathrm{D}\rangle$ ($\sim$~40\% - B.~Pagel 2003, private
communication) this can not be used either to select between the two
classes of models.

For both $^4$He and D, if $\eta$ is increased beyond $\eta$=10 the further
change in $Y_p$ and $Y_\mathrm{D}$ for a given $\eta_\mathrm{m}$ is small 
compared to the above mentioned observational uncertainties. 

\subsection{CMB Anisotropy and Polarization}

As was mentioned by \citet{NN}, the presence of the baryonic
clouds in the
primordial hydrogen-helium plasma at redshift $z\sim10^3$ changes
the dynamics of  hydrogen recombination  due to the non-linear dependence of
the ionization fraction $x_\mathrm{e}$ on
the baryon density. During the period of
recombination diffusion of baryons from inner to outer regions of the
clouds can suppress any small scale irregularities inside the
clouds. The natural length scale of this process is of the order of the Jeans
length  $R_\mathrm{J} \sim c_\mathrm{s} t_{\mathrm{rec}}$,
where $c_\mathrm{s}$ is the baryonic sound speed
and $t_{\mathrm{rec}}$ is the corresponding time when the plasma became
transparent for the CMB radiation \citep{LYSN,NN}.
For  adiabatic perturbations at scales
$M\gg  10^{13}M_{\odot}$ (which are the sources of the Doppler peaks
in the CMB anisotropy and polarization power spectra) the evolution
during the period of recombination depends not on the ionization
fraction inside or outside the clouds, but rather on the mean value of the
ionization fraction at the scales of adiabatic perturbations.
This mean ionization fraction does not correspond to
the ionization fraction for the mean value of the baryonic density due
to non-linear effects.

For the cloudy baryonic model we introduce the mean value of the
electron density
\begin{equation}
\langle n_\mathrm{e} \rangle= n_{\mathrm{e,in}}f + n_{\mathrm{e,out}}(1-f),
\end{equation}
where $n_{\mathrm{e,in}}$ and $n_{\mathrm{e,out}}$ are the number
densities of free electrons
inside and outside the
clouds and
\begin{eqnarray}
n_{\mathrm{e,in}} & =& x_{\mathrm{e,in}}
\left( 1-Y_{\mathrm{He,in}}\right) n_{\mathrm{b,in}},\nonumber
 \\
n_{\mathrm{e,out}}& =& x_{\mathrm{e,out}}
\left( 1-Y_{\mathrm{He,out}} \right) n_{\mathrm{b,out}},
\end{eqnarray}
where $x_{\mathrm{e,in}}$, $x_{\mathrm{e,out}}$,
$Y_{\mathrm{He,in}}$ and $Y_{\mathrm{He,out}}$ are the
ionization fractions and helium mass fractions for the inner and outer
regions, and the nucleon number density is denoted $n_\mathrm{b}$.
 Note that by definition $x_\mathrm{e}=n_\mathrm{e}/n_\mathrm{H}$,
where $n_\mathrm{H}$
is the number  density of neutral and ionized hydrogen.
Let us introduce the mean value  of the ionization fraction $\langle
x_\mathrm{e} \rangle$,
\begin{equation}
\langle x_\mathrm{e} \rangle=\frac{\langle n_\mathrm{e} \rangle}
{\langle n_\mathrm{b} \rangle
}\left(1-\langle{Y_{\mathrm{He}}\rangle}\right)^{-1},
\end{equation}
then
\begin{equation}
\langle x_\mathrm{e} \rangle =x_{\mathrm{e,in}}G_{\mathrm{in}}+
x_{\mathrm{e,out}}G_{\mathrm{out}},
\end{equation}
where
\begin{eqnarray}
G_{\mathrm{in}}= \frac{\xi f}{1+f(\xi -1)}
\left(\frac{1-Y_{p,\mathrm{in}}}{1-\langle Y_{p} \rangle} \right),
\nonumber  \\
G_{\mathrm{out}} = \frac{1- f}{1+f(\xi -1)}
\left(\frac{1-Y_{p,\mathrm{out}}}{1-\langle Y_{p} \rangle} \right),
\end{eqnarray}
and $\langle Y_{p}\rangle$ denotes the mean mass fraction of helium.

\begin{figure}
\epsfxsize=\columnwidth
\epsfbox{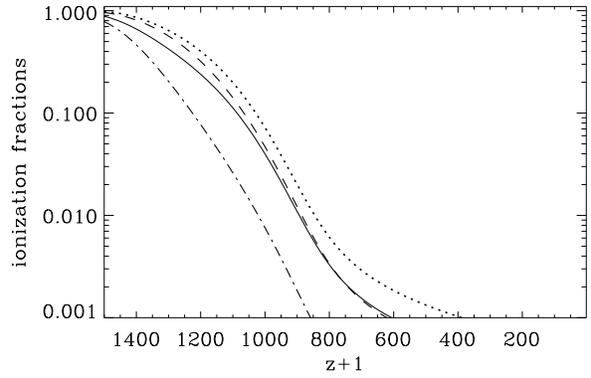}
\caption{The rate of ionization for the clumpy baryonic  model.
 The dash-dotted line corresponds to the ionization
fraction for the inner regions. The dotted line is the ionization
fraction for the
outer regions. The solid line is the mean
fraction of ionization. The dashed line is the ionization fraction
for the model with
$\Omega_{\mathrm{b}}h^2=0.022$ without clouds.}
\label{fig3}
\end{figure}

In Fig.\ \ref{fig3} we plot  the functions
$x_{\mathrm{e,in}}$,
$x_{\mathrm{e,out}}$, and $\langle x_\mathrm{e} \rangle$
for
$\langle \Omega_\mathrm{b} h^2\rangle=0.022$ , $h=0.7$, $\eta=11$, and $f=0.1$,
and cosmological parameters $\Omega_\mathrm{K}=0 $ (the curvature
parameter), $\Omega_{\Lambda}=0.7$
(vacuum density) and $\Omega_{\mathrm{WDM}}h^2=0.125$ (WDM density). A
modified version of the RECFAST code was used for the calculations.

As one
can see from this figure the fractions of ionization  $\langle x_\mathrm{e}
\rangle$, $x_{\mathrm{e,in}}$ and $x_{\mathrm{e,out}}$ have different
shapes, and
different asymptotic behaviours at ``low redshifts''.
In the range $700 \la z \la 1500$ recombination is somewhat accelerated and 
for $z \la 700$ slightly delayed compared to ``standard'' recombination.

\begin{figure}
\epsfxsize=\columnwidth
\epsfbox{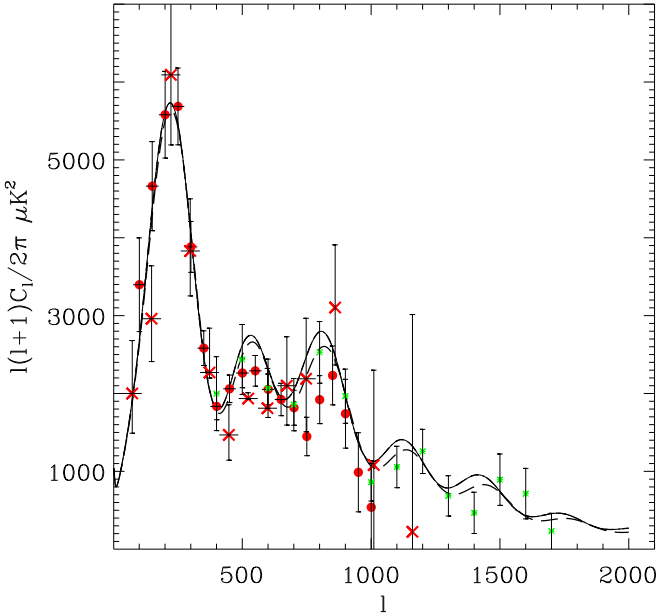}
\caption{The CMB power spectra for the WDM models. The solid line
corresponds to
the standard model, the
 dashed line to the cloudy baryonic WDM model.}
\label{fig4}
\end{figure}

In order to compare the CMB anisotropy and polarization power spectrum in the
cloudy baryonic model with
the standard one (without clouds) we 
modify the CMBFAST code \citep{SZ}, by taking into
account the more complicated
ionization history of the plasma. For the comparison we assume in both cases
(late) full reionization corresponding to
a value of the optical depth of $\tau_\mathrm{r}=0.1$.
$\tau_\mathrm{r}$ is related to the redshift of full reionization
$z_\mathrm{r}$ as
\[
 z_{\mathrm{r}}=13.6\left(\frac{\tau_\mathrm{r}}{0.1}\right)^{2/3}
 \left(\frac{1-\langle
Y_p\rangle}{0.76}\right)^{-2/3}\times
\]
\begin{equation}
\left(\frac{\langle\Omega_\mathrm{b}
h^2\rangle}{0.022}\right)^{-2/3}\left(\frac{\Omega_{\mathrm{WDM}}
h^2}{0.125}\right)^{1/3},
\label{eq7a}
\end{equation}
so $\tau_\mathrm{r}=0.1$ corresponds to a redshift of
reionization of $z_{\mathrm{r}}\simeq
14$ (the following results do not change in any significant way if 
$\tau_\mathrm{r}$=0.05 or 0.15 is adopted instead).
In Fig.\ \ref{fig4}
we plot the CMB anisotropy power
spectra for the standard and cloudy
WDM models together with the latest BOOMERANG, MAXIMA-1, and CBI
observational data (the very recent WMAP data are essentially in complete
agreement with the above data, Spergel et al. 2003). 
As can be seen from the figure
small deviations of the power spectrum appear, but both models are in
agreement with the data of the CMB experiments (see also Table 1).

\begin{figure}
\epsfxsize=\columnwidth
\epsfbox{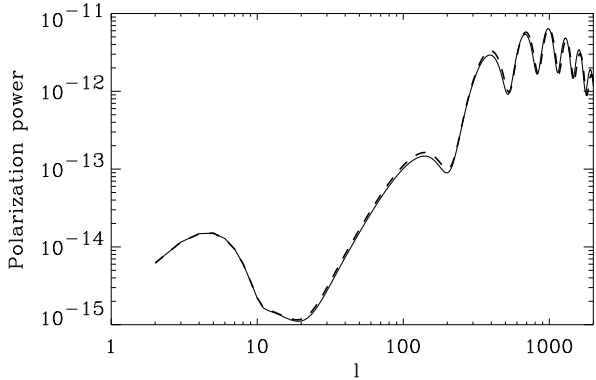}
\caption{The CMB polarization  power spectra for the WDM models. The solid line
corresponds to the standard model,
the dashed line to the cloudy baryonic WDM model.}
\label{fig5}
\end{figure}

To illustrate the effect of baryonic clouds in the WDM model and to compare 
with the standard WDM model,
taking into account the
expected sensitivity of the PLANCK experiment, we describe
the deviations of the anisotropy and polarization power spectra
in terms of  the functions
\[
D^a(l)=2\left[C^a_b(l)-C^a_s(l)\right]/\left[C^a_b(l)+
C^a_s(l)\right]\,,
\]
\be
D^p(l)=2\left[C^p_b(l)-C^p_s(l)\right]/\left[C^p_b(l)+
C^p_s(l)\right]\,,
\label{eqd}
\ee
where $C^a_b(l)$ and $C^a_s(l)$ are the CMB anisotropy power spectra for the
baryonic clouds and smooth baryon models, respectively, and $C^p_b(l)$ and 
$C^p_s(l)$ are the corresponding polarization power spectra. 
Obviously, it is necessary to compare
the functions $ D^a(l)$ and  $ D^p(l)$
with the errors of the
$C(l)$ extraction for the PLANCK mission.

\begin{figure}
\epsfxsize=\columnwidth
\epsfbox{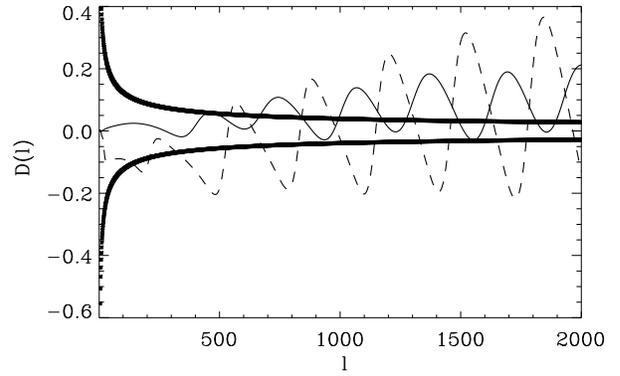}
\caption{$D^a(l)$ and  $D^p(l)$for the CMB anisotropy and polarization  power
spectra in WDM models.
The solid line corresponds to anisotropy,
the dash line is the polarization. The thick solid line corresponds to the
cosmic variance effect. }
\label{fig6}
\end{figure}

Assuming that systematic effects can be essentially removed, the uncertainty 
should be close to the cosmic variance limit
\be
\frac{\Delta C(l)}{C(l}\simeq \frac{1}{\sqrt{f_{\mathrm{sky}}(l+
\frac{1}{2})}}\left[1+w^{-1}C^{-1}(l)W^{-2}_l\right],
\label{c}
\ee
\[
w=(\sigma_{p}\theta_{\mathrm{FWHM}})^{-2},\quad W_l\simeq \exp\left[-
\frac{l(l+1)}{2l^2_s}\right],\quad f_{\mathrm{sky}}\simeq 0.65\,.
\]
Here $f_{\mathrm{sky}}$ is the sky coverage during the first
year of observations, $\sigma_{p}$ is the sensitivity per resolution
element $\theta_{\mathrm{FWHM}}\times\theta_{\mathrm{FWHM}}$
and $l_s=\sqrt{8\ln2}
\theta^{-1}_{\mathrm{FWHM}}$.
In Fig.\ \ref{fig6}
we plot the  functions $D^a(l)$ and  $D^p(l)$ for $2< l < 2000$ and show
the uncertainty due to cosmic variance.

For example, for the HFI PLANCK 214 GHz  channel $\theta_{\mathrm{FWHM}}$
is 5.5 arcmin and the errors at $l\le 2000$ are dominated by the first
term in Eq.~(\ref{c}). As one can see from Fig.\ \ref{fig6},
at $l\sim 1000-1500$
the peak-to-peak amplitudes are $\sim 10-20\%$, while $\Delta
C(l)/C(l)\simeq 5-7\%$. This means that such signals in the
anisotropy power spectrum can be detected by the PLANCK mission,
if the systematic effects can be accounted for to better than 
$\sim$~$\Delta C(l)/C(l) \simeq [f_{sky}(l+1/2)]^{-1/2}$.

\begin{table*}
\begin{minipage}{\textwidth}
\centering
\caption{CMB power spectrum $\chi^2$ parameters for the WDM and WDMB models.}
\begin{tabular}{lcc ccc }
\hline
  Obs/Mod &BOOM&MAXIMA& CBIM1 & CBIM2 & VSA\cr
\hline
  Dof     & 12 & 10   &  7    &    7  & 20\cr
\hline
 Standard model (WDM) &11.2 &14.1 & 3.20 & 7.15 & 7.08\cr
 Cloudy model (WDMB)  &8.69 &11.3 & 1.60 & 5.62 & 6.59\cr
 \hline
\end{tabular}
\end{minipage}
\label{tbl2}
\end{table*}

As one can see from Fig.\ \ref{fig6}
the polarization power spectra are even more
sensitive
to distortions of the ionization history of the Universe in the WDM model with
baryonic clouds.
Recently the DASI experiment \citep{Kov} has detected CMB
polarization without a significant presence of polarized
foregrounds. This implies that the PLANCK mission potentially will be able 
to measure the polarization of the CMB with
unprecedented accuracy, probably close to the cosmic variance limit,
which means
that peculiarities of the polarization power spectrum caused by 
baryonic clouds could be observationally detected, if present, during the 
coming years.

\section{Disc Galaxy Formation with Baryonic Clumps}

The uneven distribution of baryons in this model could of course have
consequences for the formation of disc galaxies.
The baryonic clumps are additional structures,
which would not be present in ``normal'' galaxy formation scenarios. They
could possibly disturb the assembly of discs, similar to the angular
momentum problem in CDM cosmologies
(see, e.g., Navarro \& Benz 1991, Navarro \& White 1994, Navarro \& Steinmetz
1997, Bryan 2002, Sommer-Larsen, G\"otz \& Portinari 2003). 
For CDM, the early formation
of small haloes leads to many merging events, during which the subclumps,
which later form the central disc, can loose energy and orbital
angular momentum by dynamical friction, resulting in discs with too
little specific angular momentum. The presence of baryonic clumps
may aggravate this problem, and may even effectively prevent the formation
of galactic discs. To investigate this, we ran a series of TreeSPH
simulations of galaxy formation, described in this section,
with and without such clumps.

\subsection{The Code}

We use the gridless Lagrangian N-body and Smoothed Particle Hydrodynamics
code TreeSPH described in \citet{Gslgv}, which is modelled after that of
\citet{Ghk}.

Radiative gas cooling is included in the simulations \citep{Gsd}. Some
are run with a primordial abundance cooling function, in which case radiative
heating due to a metagalactic UV field is included. This homogeneous and
isotropic, redshift-dependent UV background corresponds to the
UV field produced by AGNs and young galaxies. It is modelled after
\citet{Ghm} and switches on at a redshift of $z=6$. Its effects
on the cooling function are included as discussed e.g.\ in \citet{Gvhsl}.
Other simulations were run with an about $1/3$ solar metallicity cooling
function
(more precisely with $\FeH = -0.5$). This is the metal abundance of the
intracluster medium and can probably be considered a reasonable upper limit
to the metal abundance of disc forming gas. In this case, the metagalactic
UV field is not included, since it does not play a major role because of
the very high cooling efficiency due to the metals. Furthermore, all
simulations
incorporate inverse Compton cooling, which is also explicitly
redshift-dependent.

For the gas dynamics, the smoothing length of each SPH particle is adjusted
to keep the number of neighbors close to 50, and the shear-free Balsara
viscosity \citep{Gb} is used, instead of the standard Monaghan-Gingold
viscosity \citep{Gmg}.

\subsection{The Star Formation Recipe}

Star formation is accounted for by converting SPH particles into
star particles, keeping the total number of these particles
constant. We assume that star formation sets in when the hydrogen number
density exceeds a threshold of $n_\mathrm{H,c}$= $0.01 \cm^{-3}$. This may
seem quite low, but is
still large enough to ensure that the gas has cooled below a critical
temperature of $T \sim 10^4 \Kelvin$, at which the radiative cooling
function is effectively truncated. In addition to the density criterion,
the divergence of the velocity field has to be negative in the star forming
region. As will be seen later, this second condition is very important
for simulations with WDMB.

Star formation occurs then on a time scale
\[
\ts=\frac{\tdyn}{\eps}=\frac{1}{\sqrt{4\pi G\rhogas}}\frac{1}{\eps},
\]
governed by the dynamical time scale $\tdyn$ and a star formation
efficiency (SFE) $\eps$. The SFE is quite small in the Galactic disc
at present, at most a few percent (e.g. Silk 1997). Therefore,
most simulations are run with a SFE $\eps=0.01$. But to account for the
possibility of more efficient star formation in the early universe,
we also use $\eps=0.1$ and $\eps=1$ for some of the simulations.
As is customary, the actual point in time,
when a SPH particle is converted to a star particle, is then determined
probabilistically, based upon its star formation time scale.

The SFE is kept constant in time, which is of course unrealistic for the
cases with high SFE $\eps=0.1$ and $\eps=1$. The high-SFE simulations
were primarily run to see what effect the SFE has on the onset of star
formation at high redshifts.

Feedback is not included in the simulations. Even though it may
be important in the early stages of galaxy formation (e.g. Sommer-Larsen,
G\"otz \& Portinari 2003),
it is probably not at later times. In the warm dark matter (WDM) cosmological
scenario, which is used here (see below), the angular momentum problem,
as well as other problems on galactic scales related to CDM,
can be significantly reduced without invoking feedback \citep{Gsld}. In
any case, the absence of feedback in the simulations can be regarded as
a conservative approach, since its inclusion would possibly result in even 
more realistic galactic discs \citep{Gslgp}.

\subsection{Cosmological Parameters and Initial Conditions}

\begin{table*}
\begin{minipage}{\textwidth}
\centering
\caption{Physical properties of the two selected disc galaxies in the
different simulation runs. A `0' in the
column `\# of clumps' means that the simulation was run with evenly
distributed baryonic matter, otherwise the number of clumps in the
initial galaxy forming region is given. $\FeH$ is the metallicity,
with `prim.' denoting the primordial value, $\eps$ the star formation
efficiency, $\Mstar$ and $\Mcg$ the masses in stars and cold gas
($T< 3\cdot 10^4 \Kelvin$), respectively, $\Vc$ the circular velocity,
$\jtot$ the specific angular momentum of the disc, including
stars and cold gas, and $\jstar$ the specific angular momentum of the
stars only. The quantities in the last five columns are for
within $40\kpc$ of the center of the disc.}
\begin{tabular}{||c||c|c|l|c|r@{.}l|c|r@{}l|c||} \hline\hline
Galaxy & \# of & $\FeH$ & \multicolumn{1}{c|}{$\eps$} &
  $\Mstar$ & \multicolumn{2}{c|}{$\Mcg$} &
  $\Vc$ & \multicolumn{2}{c||}{$\jtot$} & $\jstar$ \\
 & clumps & & &
  $\left[ 10^{10} M_\odot\right]$ &
  \multicolumn{2}{c|}{$\left[ 10^{10} M_\odot\right]$} &
  $\left[ \km/\sec\right]$ &
  \multicolumn{2}{c||}{$\left[ \kpc\cdot\km/\sec\right]$} &
  $\left[ \kpc\cdot\km/\sec\right]$ \\ \hline\hline
19 &   0 & prim. & 0.01 & 4.97 & \hspace{1.1em}0&83  & 209 &  &511 & 341 \\
19 &   0 & -0.5  & 0.01 & 6.98 & 1&81  & 218 & \hspace{1.5em}1&433 & 961 \\
   \hline
19 & 248 & prim. & 0.01 & 5.41 & 0&76  & 213 &  &557 & 422 \\
19 & 248 & prim. & 0.1  & 4.64 & 0&05  & 195 &  &271 & 270 \\
19 & 248 & prim. & 1.0  & 4.35 & 0&004 & 185 &  &292 & 292 \\
19 & 248 & -0.5  & 0.01 & 7.98 & 1&27  & 229 & 1&119 & 878 \\ \hline\hline
27 &   0 & prim. & 0.01 & 3.50 & 0&53  & 179 &  &466 & 224 \\
27 &   0 & -0.5  & 0.01 & 5.17 & 1&01  & 196 &  &934 & 493 \\ \hline
27 & 177 & prim. & 0.01 & 3.65 & 0&62  & 182 &  &536 & 331 \\
27 & 177 & prim. & 0.1  & 3.70 & 0&27  & 178 &  &458 & 336 \\
27 & 177 & prim. & 1.0  & 3.34 & 0&09  & 161 &  &432 & 404 \\
27 & 177 & -0.5  & 0.01 & 6.31 & 1&13  & 217 &  &851 & 621 \\ \hline\hline
\end{tabular}
\end{minipage}
\label{tabGal}
\end{table*}

The cosmological initial conditions are based on a flat $\Lambda$WDM model
with $\OM=0.3$ and $\OL=0.7$. The Hubble parameter is chosen to be
$H_0=100h \km/\sec/\Mpc=65 \km/\sec/\Mpc$, yielding a present age of the
universe of $14.5 \Gyr$. The initial power spectrum of density fluctuations
is cluster abundance normalized to give a $\sigma_8=1.0$ at $z$=0 
(Eke, Cole \& Frenk 1996 --- adopting $\sigma_8=0.9$ does not lead to any
significant changes of the results presented in the following). 
For the WDM particle,
we use a free-streaming mass of $1.5\cdot 10^{11} h^{-1} M_\odot$,
corresponding to a free-streaming scale of $0.11 h^{-1}\Mpc$ and a
WDM particle mass of $1.2 h^{5/4} \keV$. These numbers are consistent
with the values used by \citet{Gsld} for
solving (to within a factor of two) the angular momentum problem with WDM. 
In addition,
the same sharp cut-off of the WDM power spectrum below the free-streaming
scale is used here.

In order to select interesting candidate galaxies for further study,
we ran at first a dark matter-only N-body simulation in a $10 h^{-1}\Mpc$ box
with $128^3$ particles, using the Hydra code \citep{Ghydra} and
starting at an initial redshift $\zini=19$. The usual set-up, where
particles are displaced from a regularly spaced grid, can not be used
in WDM simulations, as this leads to the formation of spurious low-mass
haloes evenly spaced along the forming filaments, which are non-physical
artifacts. (See G\"otz \& Sommer-Larsen 2002 for a detailed discussion.) 
We therefore
start from glass-like initial conditions \citep{Gw} where the particles
are irregularly distributed, but still (almost) evenly spaced.

At redshift $z=0$, haloes in the pure dark matter simulation are identified
with a friends-of-friends algorithm, where gravitationally unbound
particles are removed. The galaxies then selected for detailed simulation
with the TreeSPH code had to be sufficiently isolated (at least $1 \Mpc$
away from galaxy groups and $0.5 \Mpc$ away from larger galaxies at
$z=0$) and had not to undergo a major merger (with mass ratio more
than 1:3) since $z=1$.

To simulate the formation of the selected galaxies, baryonic matter
has to be added to the initial conditions. We aimed at obtaining
a baryonic fraction
$\fb\approx 0.1$, consistent with nucleosynthesis constraints and with observed
baryonic fractions in groups and clusters (see e.g.\ \citet{Gef}). Our
choice of $\fb$ is a bit low compared to recent estimates of $\Ob$ and $\OM$
(section~4), but this has no consequences for any of the following results.

For the simulations with baryonic clumps, we choose the most extreme
situation where all of the baryonic matter is initially contained inside
spherical clumps, whereas the dark matter is exclusively found outside of
them. That way, we maximize the effect, which the clumps have on galaxy
formation. The easiest way to achieve the desired baryonic fraction
$\fb\approx 0.1$ is then to remove $10\%$ of the original $128^3$ dark matter
particles from spherical subregions of the cosmological simulation,
and fill these regions with gas particles. To get (approximately)
the same number of particles as in the simulations, where the baryonic
matter is distributed evenly, we use $128^3$ SPH particles for that purpose,
such that each has a mass $\fb$ times that of a dark matter particle.
We therefore get in the clumps a $10\times$ higher resolution
(in terms of particle number density) than in
the surroundings, where only dark matter is found, and can thus study the
interaction of the clumps in great detail.

Because we are constrained by the fact, that there has to be an integer
number of clumps, each containing an integer number of SPH particles,
we ended up with placing $128^2=16384$ baryonic clumps into the box,
each with $128$ baryonic particles and replacing $13$ dark matter particles.
This gives a baryonic fraction $\fb=0.102$, very close to the desired
value. The initial conditions with the clumps contain then
$\left( 1-\fb\right)\cdot 128^3 = 1884160$ dark matter particles
(with a mass of $\mDM=4.0\cdot 10^7 h^{-1}M_\odot$ for the $10 h^{-1}\Mpc$ box)
and $128^3$ SPH particles (with a mass of $\mb=4.0\cdot 10^6 h^{-1}M_\odot$).
The mass and radius of a clump are $\Mclump = 5.2\cdot 10^8 h^{-1}M_\odot$
and $\Rclump = 0.11 h^{-1}\Mpc$, respectively, with an average separation
of $\dclump = 0.39 h^{-1}\Mpc$ (all length scales given are in comoving 
coordinates). Our choice of clump mass is motivated by the desire to have
$\Mclump \ll M_{\mathrm{fs}}$, but still large enough to be resolved in the
simulations. Any other choice of $\Mclump$ would 
lead to very similar results, provided $\Mclump \ll M_{\mathrm{fs}}$ --- see 
below. 

The clumps are distributed irregularly throughout the simulation volume,
with their centers forming a glass-like structure. The $13$ closest
dark matter particles to a clump center are removed, and the resulting
void is filled with a sphere containing $128$ SPH particles,
also with a glass-like distribution, which furthermore is oriented at
random. This procedure assures that the initial density field is not
disturbed, and that neighboring clumps do not overlap initially. (Random
placement of the clump centers would result in many overlapping clumps as
their average separation is not much larger than their size,
$\dclump \approx 3.5\Rclump$.) 

For the control simulations with evenly distributed baryonic matter,
one SPH per dark matter particle was added within one SPH softening length
\citep{Gsld} and with a mass $\fb$ times
the original dark matter particle mass. Since no dark matter particles
were removed, their masses were reduced to $\left( 1-\fb\right)$ the
original value. Thus $\mb$ is the same as in the simulations with clumps,
whereas now $\mDM=3.6\cdot 10^7 h^{-1}M_\odot$.

From these initial conditions, which now contain SPH particles, regions
were cut out corresponding to the subvolumes, from which the galaxies
of interest form in the pure dark matter simulation, as identified by
tracing their particles back to the initial conditions.
(Care is taken to make sure that  no clumps are intersected by the region
boundaries in simulations with baryonic clumps.) Only inside these areas
are the gas particles included, whereas for the surrounding region the
original pure dark matter initial conditions were used, with the dark
matter particles increasingly coarsely resampled with increasing distance
from the galaxy forming subvolume. That way, the formation of individual
galaxies can be followed in detail within the correct cosmological
setting, see e.g.\ \citet{Gnw,Ggelsl,Gtc,Gslgp}.

Initially, the SPH particles are assigned a thermal energy corresponding
to a temperature $T\approx 55\Kelvin$. Gravitational interactions are
softened according to the prescription of \citet{Ghk}.
The softening lengths are $1.3h^{-1}\kpc$ for gas and star particles,
$2.9h^{-1}\kpc$ for the dark matter particles inside the galaxy
forming region in simulations with clumps, and $2.8h^{-1}\kpc$ for such dark
matter particles in runs with evenly distributed baryonic matter. The
softening lengths of the dark matter particles in the coarsely resampled
surrounding are correspondingly larger.

\subsection{The Simulations}

We selected two dark matter haloes (numbers 19 and 27) from the
cosmological N-body simulation for further study with the TreeSPH code.
The haloes have characteristic
circular velocities comparable to or a little below that of
the Milky Way. These galaxies were chosen according to the criteria
laid out above, and because they show distinct disc-like morphologies
and exponential surface density profiles in the TreeSPH-runs with
evenly distributed baryonic matter.
For each of the haloes, six simulations with different
combinations of cooling function metallicity, star formation efficiency, and
presence or absence of baryonic clumps, were run, as can be seen in
Table \ref{tabGal}. The simulations of the larger halo, \#19, contain
about 34000 dark matter and 22000 (without clumps) or 32000 (with clumps)
SPH particles initially, whereas the runs for halo \#27 consist of about
26000 dark matter and 14000 (without clumps) or 23000 (with clumps)
gas particles at the start. The
significantly higher number of SPH particles in simulations with baryonic
clumps is due to the fact that their initial regions have to be extended
to avoid that clumps are intersected by the boundaries of the galaxy-forming
subvolumes.

\subsection{Results}

\subsubsection{Properties at $z=0$}

\begin{figure}
\epsfxsize=\columnwidth \epsfbox{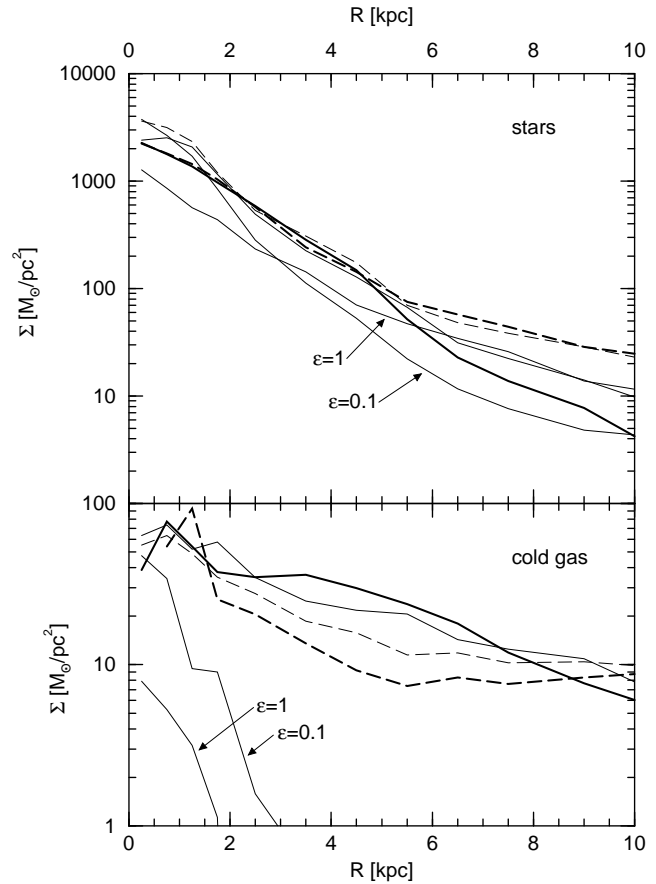}
\caption{Surface density profiles for Milky Way like disc galaxy \#19
in stars (top) and cold gas with $T< 3\cdot 10^4 \Kelvin$ (bottom) at $z=0$.
Thick lines denote runs with evenly distributed baryonic matter, and thin lines
those with baryonic clumps. Simulations with primordial abundance are
shown as solid lines, and the 1/3 solar metallicity cases ($\FeH=-0.5$) by
dashed lines. The SFE is $\eps=0.01$, unless otherwise noted. The scale
lengths for the stellar discs lie between $1 \kpc$ and $1.5 \kpc$.}
\label{figsurfd19}
\end{figure}
\begin{figure}
\epsfxsize=\columnwidth \epsfbox{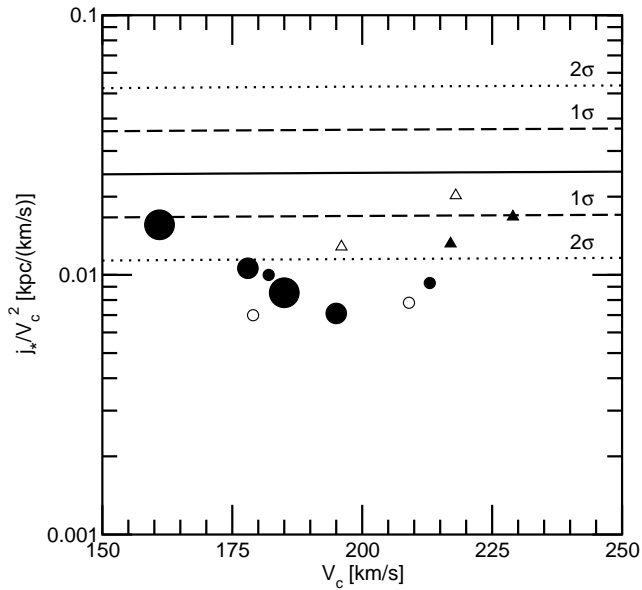}
\caption{The normalized specific angular momenta in stars
$\jstar/\Vc^2$ at $z=0$ of the disc galaxies formed in the simulations. Small
symbols are for runs with a SFE $\eps=0.01$, medium sized ones have $\eps=0.1$,
and the large ones $\eps=1$. The round symbols denote primordial abundance,
and the triangles a metallicity of $\FeH=-0.5$. Open symbols are used for
the control simulations with evenly distributed baryonic matter, whereas
filled symbols are for runs with baryonic clumps. The solid line shows
the median value from observational data on disc galaxies \citep{Gbyun},
obtained as explained in \citet{Gsld}, but using $H_0=65 \km/\sec/\Mpc$,
the dashed and dotted lines bracket the $1\sigma$ and $2\sigma$ intervals
around the median.}
\label{figjtot}
\end{figure}
In all of the TreeSPH-simulations, the galaxies show clear disc-like
morphologies and kinematics,
with the bulk of the stars on approximately circular orbits in the disc.
Most of the remaining
stars are found in an inner, bulge-like component and finally a small
fraction in a dynamically insignificant stellar halo surrounding the disc.
The disc galaxies formed in our simulations are thus qualitatively very
similar to observed disc galaxies, like the Milky Way, irrespective of
the presence or absence of baryonic clumps initially,
and of the assumed values for the IGM metallicity and SFE.

All of the discs have approximately exponential surface density profiles
in stars and cold gas out to four scale lengths. As an example, the
profiles for the different simulations of galaxy \#19 are shown in
Fig.\ \ref{figsurfd19}. The exponential scale lengths for the stellar
discs lie between $1 \kpc$ and $1.5 \kpc$. The runs with 1/3 solar metallicity
(dashed lines in Fig.\ \ref{figsurfd19}) show somewhat flatter gas and 
stellar profiles
and a lower density in cold gas, since the more effective cooling leads
to a higher conversion rate of gas into stars. The simulations with a
higher SFE (as marked in Fig.\ \ref{figsurfd19}) have, as one would expect,
much depleted gas discs, and lower surface densities in stars, since their
star formation rates peak at higher redshifts (see below). Thus more
stars are formed initially in the halo, and less gas is left to form
the disc stars.

The scale lengths of the stellar discs (see Fig.\ \ref{figsurfd19})
are somewhat low compared to observed values --- thus the angular momentum
problem has not been completely overcome.
This can also be seen by looking at the normalized
specific angular momenta in stars $\jnormstar=\jstar/\Vc^2$ plotted
in Fig.\ \ref{figjtot} for the two discs in all the different simulation
runs. As argued by \citet{Gslgv} one expects $\jnormstar$ to be almost
independent of $\Vc$ on both theoretical and observational grounds. On
average, the normalized specific angular momenta of our discs lie only
about a factor of two below the median observed value, as obtained by
\citet{Gsld}. This compares well with other attempts to solve the
angular-momentum problem, e.g.\ the WDM models of \citet{Gsld} and CDM
scenarios with feedback \citep{Gslgp}, which also come within a factor
of two to three of the observed specific angular momenta.

In Table \ref{tabGal} and Figs.\ \ref{figsurfd19} and \ref{figjtot}, we have
included the results for the simulations with high SFE $\eps=0.1$ and
$\eps=1$ at $z=0$, even though these runs were initially carried out
only to study the effect, a possible high SFE at large redshifts would
have on early star formation. As expected, these simulations produce discs
which have an unrealistically low cold gas fraction, indicating that the
SFE can not stay constant at these large values. In spite of that, their
specific angular momenta lie in the same range as that of the other runs
(see Fig.\ \ref{figjtot}), indicating that SFE is not an important parameter
in WDM models with baryonic clumps for solving the angular momentum problem. 

\subsubsection{The High-Redshift Evolution}

\begin{figure}
\epsfxsize=\columnwidth \epsfbox{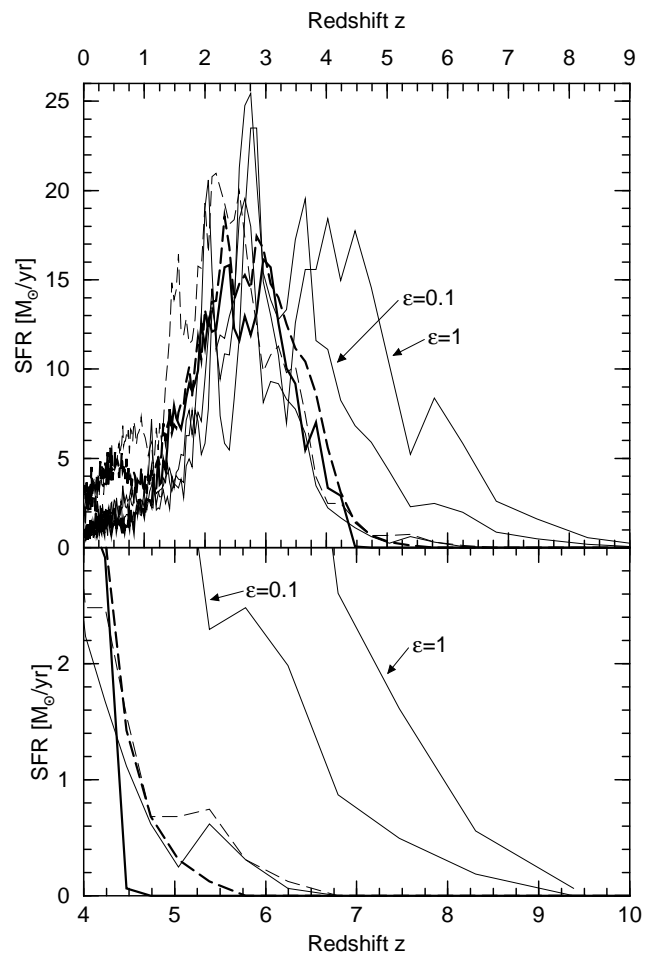}
\caption{Star formation rates in $M_\odot/\yr$ for the different simulations
of galaxy \#19, with a magnification of the high-redshift tail at the
bottom. Line styles are as in Fig.\ \ref{figsurfd19}.}
\label{figsfr19}
\end{figure}
\begin{figure}
\epsfxsize=\columnwidth \epsfbox{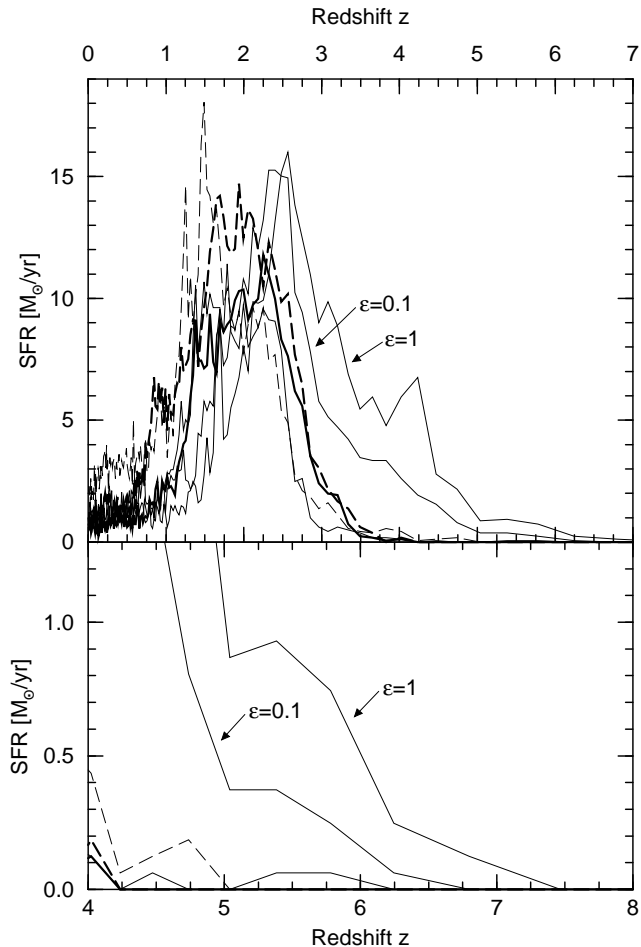}
\caption{Same as Fig.\ \ref{figsfr19}, but now for galaxy \#27.}
\label{figsfr27}
\end{figure}
The present-day properties of disc galaxies are thus unaffected by the
presence or absence of baryonic clumps. In both cases do we find discs at
redshift $z=0$ with the same kinds of profiles and specific angular
momenta. The reason for this lies in the behavior of the clumps at
the beginning of structure formation.

When the first sheets and filaments form at high redshifts, the baryonic clumps
get rapidly distorted into ellipsoidal shapes by the
tidal field of these structures. The clumps get `gently' incorporated
into the filaments, and
there are no shocks occurring during this, or when the clumps eventually
physically touch. 
After $1.0$--$1.5$ Gyr ($z$=4--5), the filaments in the simulations with 
baryonic
clumps are almost evenly filled with baryonic matter, after which the
same processes of galaxy formation take place as in the ordinary WDM
scenario with evenly distributed baryonic matter (The clumps do though
survive much longer in the voids, but this is not of interest here, since
the early galaxy/star formation in the simulations takes place in
the filaments). During filament formation, 
also the dark matter becomes distributed smoothly in the filaments. This
general behaviour strongly suggests that the main results of the simulations
do not depend of the specific choice of clump mass, as previously mentioned.

But there is a significant difference between the clump and no-clump models
in terms of the early star formation history. 
Figs.\ \ref{figsfr19} and \ref{figsfr27}
show the star formation rates for the two selected galaxies in all the
model runs. In the simulations with baryonic clumps, the first stars
form earlier than in those without them. This is not surprising,
since the baryonic matter in the models with clumps starts out with a
higher density. If one only applied a density criterion as the star
formation recipe, one could get star formation started very early in
WDMB. But the initial gravitational compression and deformation of the
clumps heats the baryonic material, and afterwards a balance between Hubble
expansion and tidal gravitational squeezing is achieved for some time,
keeping the baryonic density in the clumps approximately constant. 
Only after turn-around
has the baryonic material in the clumps a chance to cool efficiently and
to start forming stars. This is why we have included the condition, that
the gas velocity field needs to be compressive in star
forming regions, into our star formation recipe.

The first stars form at an epoch $z_\mathrm{*,i}$ which is of order a
period $\ts$ after the first region with a compressive gas velocity field
and $n_\mathrm{H}>n_\mathrm{H,c}$ appears. For the standard WDM simulations
with $\epsilon=0.01$ $z_\mathrm{*,i}\simeq4.5$; cf. Figures 9 and 10 (the
[Fe/H]=-0.5 cooling function simulations are clearly inappropriate when
discussing the formation of the {\it first} stars). The similar WDMB
simulations have $z_\mathrm{*,i}\simeq6.5$. The WDMB simulations with
higher SFEs ($\epsilon=0.1$ and 1) have $z_\mathrm{*,i}$=7--9.5. We did not
run standard WDM simulations with $\epsilon=0.1$ and 1, but based on the
$\epsilon=0.01$ simulations we estimate that even for such high SFEs
$z_\mathrm{*,i}\la5.5$. Hence going to the WDMB model brings the warm
dark matter model in much better agreement with observations of high-$z$
galaxies (e.g. Hu et al. 2002 --- $z$=6.56) and QSOs (e.g. Fan et al. 2001 ---
several with $z>6$). In fact for the high SFE simulations there is even
marginal agreement with the value of the redshift of reionization of the 
Universe
$z_\mathrm{re}\ga10$ recently claimed by the WMAP team \citep{S.03}, provided
reionization happens shortly after the onset of star formation --- but see
below. We note that \citet{Gslgp} find that the SFE likely was considerably 
higher during the early phases of galaxy formation than at present. For
$z\ga10$ the gas velocity field is everywhere expansive, so it will not
be possible to push $z_\mathrm{*,i}$ beyond $\approx$10.

\section{Summary and Conclusions}

We have investigated the implications of strongly inhomogenous, primordial 
baryon distribution on sub-galactic scales 
(10$M_{\odot} \ll M \ll 10^{11}M_{\odot}$) for Big Bang Nucleosynthesis, 
CMB anisotropies and galaxy formation in the context of the warm dark matter
model. 
Big Bang Nucleosynthesis is only slightly changed relative to SBBN, but
the change in recombination history at $z~\sim~1500-700$ relative to
``standard'' theory leads to differences in the anisotropy and polarization
power spectra, which should be detectable by the Planck satellite provided
systematic effects can be understood. We show by fully
cosmological, hydro/gravity simulations that the formation of galactic
discs is only weakly affected by going from the WDM to the WDMB scenario. 
In particular, the final disc angular momenta (at $z$=0)
are as large as for the standard case and the ``disc angular momentum
problem'' is solved to within a factor of two or better without invoking
(hypothetical) energetic feedback events.  A very desirable
difference from the standard WDM model, however, is that the on-set of star 
(and AGN) formation happens
earlier. For the ``optimal'' free-streaming
mass of $M_f~\sim~1.5~\cdot~10^{11}~h^{-1}~M_{\odot}$ the redshift
of formation of the first stars increases from $z_\mathrm{*,i}$=4-5 to 
$\ga$6.5, in much better
agreement with observational data on high-redshift galaxies and QSOs. The
WDMB model is, however, for this choice of free-streaming mass still
limited to $z_\mathrm{*,i}$\la10, so probing the ``dark ages'' with
JWST, ALMA etc. will offer direct tests of this theory.

\section{Acknowledgements}

We have benefited from comments by M.~Demianski, A.~Doroshkevich, 
L.~Portinari and M.~Way. This project was supported by Danmarks
Grundforskningsfond through its support for the establishment of the
Theoretical Astrophysics Center.

\label{lastpage}


\begin{thebibliography}{99}
\bibitem[\protect\citeauthoryear{Abramo \& Finelli}{2001}]{AF}
  Abramo L.R., Finelli F., 2001, Phys.\ Rev.\ D, 64, 083513
\bibitem[\protect\citeauthoryear{Affeck \& Dine}{1985}]{AD}
  Affeck I., Dine M., 1985, Nucl.\ Phys.\ B, 249, 361
\bibitem[\protect\citeauthoryear{Balsara}{1995}]{Gb}
  Balsara D.S., 1995, J.\ Comp.\ Phys., 121, 357
\bibitem[\protect\citeauthoryear{Bartolo, Matarrese \& Rioto}{2001}]{BMR}
  Bartolo N., Matarrese S., Rioto A., 2001, Phys.\ Rev.\ D, 64, 123504
\bibitem[\protect\citeauthoryear{Bode, Ostriker \& Turok}{2001}]{BOT}
  Bode, P., Ostriker, J.P., Turok, N., 2001, ApJ, 556, 93
\bibitem[\protect\citeauthoryear{Bond \& Crittenden}{2001}]{BC}
  Bond, R. J.,Crittenden, R., 2001, in Crittenden, R., Turok, N., eds.,
  Structure Formation in the Universe. Kluwer, Dordrecht, p. 241
\bibitem[\protect\citeauthoryear{Bryan}{1999}]{Gb99}
  Bryan G.L., 1999, Computing in Science and Engineering, 1:2, 46
\bibitem[\protect\citeauthoryear{Bryan}{2002}]{Gb02}
  Bryan G.L., 2002, in preparation
\bibitem[\protect\citeauthoryear{Bucher, Moodley \& Turok}{2001}]{BMT}
  Bucher M., Moodley K., Turok N., 2001, Phys.\ Rev.\ Lett., 87, 191301
\bibitem[\protect\citeauthoryear{Byun}{1992}]{Gbyun}
  Byun Y.-I., 1992, PhD thesis, The Australian National
  University
\bibitem[\protect\citeauthoryear{Carlberg et al.}{1996}]{Ca}
  Carlberg, R.G., et al., 1996, ApJ, 462, 32
\bibitem[\protect\citeauthoryear{Couchman, Thomas \& Pearce}{1995}]{Ghydra}
  Couchman H.M.P., Thomas P.A., Pearce F.R., 1995, ApJ, 452, 797
\bibitem[\protect\citeauthoryear{de Bernardis et al.}{2000}]{dB}
  de Bernardis P. et al., 2000, Nature, 404, 955
\bibitem[\protect\citeauthoryear{Dolgov \& Silk}{1993}]{DS}
  Dolgov A., Silk J., 1993, Phys.\ Rev.\ D, 47, 2619
\bibitem[\protect\citeauthoryear{Demianski \& Doroshkevich}{2003}]{DD}
  Demianski, M., Doroshkevich A.G., 2003, ApJ, in press (astro-ph/0304484)
\bibitem[\protect\citeauthoryear{Doroshkevich et al.}{2003}]{DNNN}
  Doroshkevich A.G., Naselsky I.P., Naselsky P.D., Novikov I.D., 2003,
  ApJ, 586, 709
\bibitem[\protect\citeauthoryear{Doroshkevich, Zel'dovich \& Novikov}{1967}]{DZN}
  Doroshkevich, A., Zel'dovich, Ia. B., Novikov, I. D., 1967, Sov. Astr., 11, 233
\bibitem[\protect\citeauthoryear{Eke, Cole \& Frenk}{1996}]{Gecf}
  Eke V.R., Cole S., Frenk C.S., 1996, MNRAS, 282, 263
\bibitem[\protect\citeauthoryear{Ettori \& Fabian}{1999}]{Gef}
  Ettori S., Fabian A.C., 1999, MNRAS 305, 834
\bibitem[\protect\citeauthoryear{Fan et al.}{2001}]{Fan.}
  Fan, X., et al., 2001, AJ, 122, 2833
\bibitem[\protect\citeauthoryear{Fukugita, Hogan \& Peebles}{1998}]{FHP}
  Fukugita, M. Hogan, C.J., Peebles, P.J.E., 1998, ApJ, 518, 503
\bibitem[\protect\citeauthoryear{Gelato \& Sommer-Larsen}{1999}]{Ggelsl}
  Gelato S., Sommer-Larsen J., 1999, MNRAS 303, 321
\bibitem[\protect\citeauthoryear{Gnedin \& Ostriker}{1992}]{GO}
  Gnedin, N.Iu., Ostriker, J.P., 1992, ApJ, 400, 1
\bibitem[\protect\citeauthoryear{G\"{o}tz \& Sommer-Larsen}{2002}]{Ggsl}
  G\"{o}tz M., Sommer-Larsen J., 2002, Astroph.\ Sp.\ Sc., 281, 415
\bibitem[\protect\citeauthoryear{Haardt \& Madau}{1996}]{Ghm}
  Haardt F., Madau P., 1996, ApJ, 461, 20
\bibitem[\protect\citeauthoryear{Hanany et al.}{2000}]{Han}
  Hanany S. et al., 2000, ApJ, 545, L5
\bibitem[\protect\citeauthoryear{Hernquist \& Katz}{1989}]{Ghk}
  Hernquist L., Katz N., 1989, ApJS, 70, 419
\bibitem[\protect\citeauthoryear{Hogan \& Loeb}{1993}]{HL}
  Hogan, C., Loeb, A., 1993, ApJ, 415, 63
\bibitem[\protect\citeauthoryear{Hu et al.}{2002}]{Hu.}
  Hu, E.M., et al., 2002, ApJ, 568, L75
\bibitem[\protect\citeauthoryear{Jedamzik \& Rehm}{2001}]{JR}
  Jedamzik K., Rehm J.B., 2001, Phys.\ Rev.\ D, 64, 023510
\bibitem[\protect\citeauthoryear{Klypin et al.}{1999}]{K.}
  Klypin, A., Kravtsov, A. V., Valenzuela, O., Prada, F., 1999, ApJ, 523, 32
\bibitem[\protect\citeauthoryear{Leitch et al.}{2002}]{Kov}
  Leitch, E.M., et al., 2002, Nature, 410, 763L
\bibitem[\protect\citeauthoryear{Lesgourgues \& Peloso}{2000}]{LP}
  Lesgourgues, J., Peloso, M., 2000, Phys.\ Rev.\ D, 62, 081301 
\bibitem[\protect\citeauthoryear{Liu et al.}{2001}]{LYSN}
  Liu G.-C., Yamamoto K., Sygiyama N., Nishioka H., 2001, ApJ, 547, 1
\bibitem[\protect\citeauthoryear{Malaney \& Butler}{1989}]{MB}
  Malaney R.A., Butler M.N., 1989, Phys.\ Rev.\ Lett., 62, 117
\bibitem[\protect\citeauthoryear{Mason et al.}{2003}]{Mason}
  Mason, B.S., et al., 2003, ApJ, 591, 540 
\bibitem[\protect\citeauthoryear{Monaghan \& Gingold}{1983}]{Gmg}
  Monaghan J.J., Gingold R.A., 1983, J.\ Comp.\ Phys., 52, 374
\bibitem[\protect\citeauthoryear{Moore et al.}{1999a}]{Ma.}
  Moore, B., et al., 1999a, ApJ, 524, L19
\bibitem[\protect\citeauthoryear{Moore et al.}{1999b}]{Mb.}
  Moore, B., et al., 1999b, MNRAS, 310, 1147 
\bibitem[\protect\citeauthoryear{Naselsky \& Novikov}{2002}]{NN}
  Naselsky P.D., Novikov I.D., 2002, MNRAS, 334, 137
\bibitem[\protect\citeauthoryear{Navarro \& Benz}{1991}]{Gnb}
  Navarro J.F., Benz W., 1991, ApJ, 380, 320
\bibitem[\protect\citeauthoryear{Navarro, Frenk \& White}{1995}]{Gnfw}
  Navarro J.F., Frenk C.S., White S.D.M., 1995, MNRAS, 275, 56
\bibitem[\protect\citeauthoryear{Navarro \& Steinmetz}{1997}]{Gns}
  Navarro J.F., Steinmetz M., 1997, ApJ, 478, 13
\bibitem[\protect\citeauthoryear{Navarro \& White}{1994}]{Gnw}
  Navarro J.F., White S.D.M., 1994, MNRAS, 267, 401
\bibitem[\protect\citeauthoryear{Novikov, Schmalzing \& Mukhanov}{2000}]{NSM}
  Novikov D.I., Schmalzing J., Mukhanov V.F., 2000, A\&A, 364, 17
\bibitem[\protect\citeauthoryear{Olive, Steigman \& Walker}{2000}]{OSW}
  Olive K.A., Steigman G., Walker T.P., 2000, ApJ, 333, 389
\bibitem[\protect\citeauthoryear{Peebles}{1967}]{P67}
  Peebles, P.J.E., 1967, ApJ, 147, 859
\bibitem[\protect\citeauthoryear{Peebles}{2001}]{P01}
  Peebles, P.J.E., 2001, ApJ, 557, 495
\bibitem[\protect\citeauthoryear{Peebles \& Juszkiewich}{1998}]{PJ}
  Peebles, P.J.E., Juszkiewich, R., 1998, ApJ, 509, 483
\bibitem[\protect\citeauthoryear{Peebles, Seager \& Hu}{2000}]{PSH}
  Peebles P.J.E., Seager S., Hu W., 2000, ApJ, 539, L1
\bibitem[\protect\citeauthoryear{Polarski \& Starobinsky}{1994}]{PS}
  Polarski D., Starobinsky A., 1994, Phys.\ Rev.\ D, 50, 6123
\bibitem[\protect\citeauthoryear{Riazuelo \& Langlos}{2000}]{RL}
  Riazuelo A., Langlos D., 2000, Phys.\ Rev.\ D, 62, 043504
\bibitem[\protect\citeauthoryear{Seager, Sasselov \& Scott}{2000}]{SSS}
  Seager S., Sasselov D.D., Scott D., 2000, ApJS, 128, 407
\bibitem[\protect\citeauthoryear{Seljak \& Zaldarriaga}{1996}]{SZ}
  Seljak U., Zaldarriaga M., 1996, ApJ, 469, 437
\bibitem[\protect\citeauthoryear{Silk}{1997}]{Gs97}
  Silk J., 1997, ApJ, 481, 703
\bibitem[\protect\citeauthoryear{Sommer-Larsen \& Dolgov}{2001}]{Gsld}
  Sommer-Larsen J., Dolgov A., 2001, ApJ, 551, 608
\bibitem[\protect\citeauthoryear{Sommer-Larsen, Gelato \& Vedel}{1999}]{Gslgv}
  Sommer-Larsen J., Gelato S., Vedel H., 1999, ApJ, 519, 501
\bibitem[\protect\citeauthoryear{Sommer-Larsen, G\"{o}tz \& Portinari}{2003}]{Gslgp}
  Sommer-Larsen J., G\"{o}tz M., Portinari L., 2003, ApJ, in press (astro-ph/0204366)
\bibitem[\protect\citeauthoryear{Spergel et al.}{2003}]{S.03}
  Spergel, D. N., et al., 2003, ApJ, in press (astro-ph/0302209)
\bibitem[\protect\citeauthoryear{Sutherland \& Dopita}{1993}]{Gsd}
  Sutherland R.S., Dopita M.A., 1993, ApJS, 88, 253
\bibitem[\protect\citeauthoryear{Tegmark \& Zaldarriaga}{2000}]{TZ}
  Tegmark, M., Zaldarriaga, M., 2000, Phys.\ Rev.\ Lett., 85. 2240
\bibitem[\protect\citeauthoryear{Thacker \& Couchman}{2000}]{Gtc}
  Thacker R.J., Couchman H.M.P., 2000, ApJ, 545, 728
\bibitem[\protect\citeauthoryear{Vedel, Hellsten \& Sommer-Larsen}{1994}]{Gvhsl}
  Vedel H., Hellsten U., Sommer-Larsen J., 1994, MNRAS, 271, 743
\bibitem[\protect\citeauthoryear{White}{1996}]{Gw}
  White S.D.M., 1996, in R. Schaeffer et al.\ (eds.),
  Cosmology and Large-Scale Structure: Les Houches, Session LX,
  Elsevier, Amsterdam, p. 349
\bibitem[\protect\citeauthoryear{White et al.}{2000}]{Wh2000}
  White, M., Scott, D., Pierpaoli, E., 2000, ApJ, 545, 1 
\bibitem[\protect\citeauthoryear{Yokoyama \& Sato}{1991}]{YS}
  Yokoyama J., Sato Y., 1991, ApJ, 379, 427
\bibitem[\protect\citeauthoryear{Zeldovich \& Novikov}{1983}]{ZN}
  Zeldovich, Ia. B., Novikov, I. D., 1983, Relativistic Astrophysics Vol.2,
  University of Chicago Press, Chicago, IL
\end{thebibliography}
\end{document}